\documentclass[10pt]{article}

\usepackage[utf8]{inputenc}
\usepackage[T1]{fontenc}
\usepackage{lmodern}
\usepackage[a4paper,margin=1in]{geometry}
\usepackage{setspace}
\usepackage{microtype}
\usepackage{hyperref}
\usepackage{graphicx}
\usepackage{booktabs}
\usepackage{array}
\usepackage{enumitem}
\usepackage{amsmath}
\usepackage{url}
\usepackage{float}
\usepackage{xcolor}
\usepackage{caption}
\usepackage{tabularx}
\usepackage{abstract}

\newcolumntype{Y}{>{\raggedright\arraybackslash}X}

\setlist[itemize]{leftmargin=1.5em, topsep=2pt, itemsep=2pt, parsep=0pt, partopsep=0pt}
\setlist[enumerate]{leftmargin=1.5em, topsep=2pt, itemsep=2pt, parsep=0pt, partopsep=0pt}

\hypersetup{
colorlinks=true,
linkcolor=blue,
citecolor=blue,
urlcolor=blue,
pdftitle={A Benchmarking Framework for Multimodal User Interface Toolkits},
pdfauthor={Ariton Verush},
pdfkeywords={multimodal user interfaces, multimodal toolkits, benchmarking, HCI, developer tools, logging frameworks, LLM-based interaction, evaluation methods}
}

\title{
\textbf{A Benchmarking Framework for Multimodal User Interface Toolkits}\\[0.35em]
{\large\itshape Comparing Modality Coverage, Developer Workflow, and Experimental Support}
}

\author{
Ariton Verush\\
University of Bern\\
Bern, Switzerland\\
\texttt{ariton.verush@students.unibe.ch}
}

\date{
HCI seminar draft: December 2025\\
Revised and expanded as a benchmarking framework paper: June 2026
}

\begin{document}

\maketitle

% =========================================================
% ABSTRACT
% =========================================================
\begin{abstract}
Multimodal user interfaces increasingly combine speech, gesture, vision, gaze, touch, biosignals, and other sensor data. Recent toolkits from the past five years, such as Geno, Multisensor-Pipeline (MSP), ReactGenie, and EmoSync, aim to make it easier for developers to prototype such interfaces, while older work such as WAMI shows how early web-based multimodal systems were conceived. Yet the field still lacks a systematic and reusable way to compare what these toolkits actually support, how much implementation work they offload from developers, and which evaluation strategies are appropriate for them. This paper reframes an HCI seminar draft into a benchmarking framework paper for multimodal user interface toolkits. Rather than reporting completed empirical results, it proposes a structured benchmark based on document analysis, technical comparison, and a future developer-based evaluation. The framework is organized around three dimensions: modality coverage and interaction abstraction, developer experience and workflow, and experimental and integration support. The paper illustrates the framework through five representative toolkits: Geno, MSP, ReactGenie, WAMI, and EmoSync. The contribution is a reusable benchmark template that future researchers can instantiate with empirical measurements, developer studies, and additional multimodal toolkits.
\end{abstract}


% =========================================================
% 1. INTRODUCTION
% =========================================================
\section{Introduction}

Multimodal user interfaces promise more robust and natural interaction by combining multiple input and output channels, such as speech, pointing, gaze, biosignals, touch, vision, and other modalities. Instead of forcing users to interact through a single input channel, multimodal interfaces can combine complementary signals and support richer interaction patterns. For example, speech can specify an action, pointing can identify an object, gaze can indicate attention, and sensor data can provide additional context. This makes multimodal interaction highly relevant for web applications, augmented reality, assistive technologies, intelligent environments, and emotion-aware systems \cite{Gruenstein2008WAMI,Metatla2019VoiceUISchools,Lee2024GazePointAR}.

However, multimodal interface development remains difficult in practice. Developers often need to connect heterogeneous sensors, synchronize event streams, define fusion logic, handle uncertainty, debug multiple interaction channels, and collect experimental logs. These tasks frequently require custom \textit{glue code}, project-specific workarounds, and ad-hoc evaluation scripts. As a result, even when toolkits exist, it can be difficult to compare their actual capabilities, understand their developer experience, or reuse solutions across research projects \cite{Barz2021MSP,Ledo2018ToolkitEval}.

In other areas of computing, benchmarking has played a central role in clarifying progress. Vision models, large language models, software frameworks, and computational methods are often compared through shared tasks, metrics, and evaluation protocols. For multimodal interaction toolkits, there is no equally established benchmark culture. There is no widely adopted set of tasks, criteria, or reporting practices that would allow researchers to compare frameworks such as Geno, MSP, ReactGenie, WAMI, or EmoSync in a principled way \cite{BioInfoBenchmarking2019,Ledo2018ToolkitEval}.

This paper addresses that gap by proposing a reusable benchmarking framework for multimodal user interface toolkits. The focus is on \emph{developer-facing} frameworks that support implementation, fusion, logging, and evaluation of multimodal interaction rather than on final end-user applications. The paper is intentionally framed as a benchmarking framework contribution, not as a completed empirical benchmark. Its purpose is to define what should be compared, how the comparison can be structured, and how future empirical work can instantiate the framework.

Five representative toolkits are used to illustrate the framework: Geno, Multisensor-Pipeline (MSP), ReactGenie, WAMI, and EmoSync. These systems span different eras, programming models, and design philosophies. WAMI represents earlier web-accessible multimodal architectures; Geno supports voice interaction on existing web applications; MSP focuses on multimodal and multisensor pipelines; ReactGenie represents LLM-assisted multimodal interaction generation; and EmoSync focuses on multimodal emotion-data synchronization and logging \cite{Sarmah2020Geno,Barz2021MSP,Yang2024ReactGenie,Gruenstein2008WAMI,Tong2025EmoSync}.

\subsection{Research Questions}

The proposed framework is guided by three research questions adapted from the original seminar paper:

\begin{itemize}
\item \textbf{RQ1 (Coverage).} Which modalities and recurring multimodal interaction patterns, such as speech--gesture grounding, multisensor fusion, or event-based triggering, does each toolkit support out of the box, and where do developers still need custom glue code or external services \cite{Sarmah2020Geno,Barz2021MSP,Yang2024ReactGenie,Gruenstein2008WAMI,Tong2025EmoSync}?
\item \textbf{RQ2 (Developer effort).} How can developer effort be compared when implementing similar multimodal tasks with different toolkits, and how does this relate to abstraction level, documentation quality, and debugging support \cite{Ledo2018ToolkitEval}?
\item \textbf{RQ3 (Experimental support).} Which toolkits provide built-in support for logging, synchronization, scenario replay, and experimental control \cite{Tong2025EmoSync,Barz2021MSP,Gruenstein2008WAMI}?
\end{itemize}

These questions are not treated as completed empirical claims. Instead, they define the structure of the proposed benchmark and identify what future empirical studies should measure. Instead, they define the structure of the proposed benchmark and identify what future empirical studies should measure.

\subsection{Contributions of the Paper}

This paper makes four main contributions:

\begin{itemize}
\item it reframes a seminar-based multimodal UI toolkit comparison into a one-column benchmarking framework paper suitable for public preprint release;
\item it defines three benchmark dimensions for multimodal UI toolkits: modality coverage and interaction abstraction, developer experience and workflow, and experimental/integration support;
\item it illustrates the framework through five representative toolkits: Geno, MSP, ReactGenie, WAMI, and EmoSync;
\item it provides a future evaluation protocol that can be instantiated with document analysis, prototype tasks, technical measurements, and developer studies.
\end{itemize}

The broader goal is to support clearer decision-making in multimodal UI research. Instead of asking which toolkit is universally best, the framework asks which toolkit is appropriate for which type of multimodal development task, research context, and evaluation requirement.

% =========================================================
% 2. BACKGROUND
% =========================================================
\section{Background and Related Work}

This section situates the paper within prior research on multimodal interaction toolkits, toolkit evaluation, and benchmarking methods. It first introduces the five representative toolkits, then discusses toolkit evaluation in HCI, and finally explains why benchmarking is a useful method for comparing multimodal development frameworks.

\subsection{Multimodal User Interface Toolkits}

Multimodal toolkits provide developers with abstractions for combining multiple input or output modalities. These abstractions may include speech commands, sensor streams, gesture recognition, graphical user interface context, gaze input, emotion recognition, or data-flow pipelines. Across these systems, recurring multimodal interaction patterns refer to technical combinations of modalities, fusion mechanisms, and interaction abstractions supported by the toolkit's APIs.

\textbf{Geno.}
Geno is a developer-centric framework that lets programmers \emph{teach} voice commands by demonstration on existing web applications \cite{Sarmah2020Geno}. Developers interact with the target website while providing example utterances; Geno then generates robust voice commands that integrate with the Document Object Model (DOM). It focuses on web automation with speech and GUI context, including mouse, keyboard, and DOM elements. Geno is especially relevant for developers who already work with web applications and want to augment existing interfaces with voice macros without rebuilding the entire interface from scratch.

\textbf{Multisensor-Pipeline (MSP).}
Multisensor-Pipeline is a lightweight Python framework for building multimodal and multisensor processing pipelines \cite{Barz2021MSP}. It emphasizes flexible composition of sensor nodes, stream scheduling, logging, and extensibility. MSP is closer to a signal-processing and data-flow toolkit than a classic UI widget toolkit. However, many multimodal UI prototypes rely on similar pipeline structures for fusing sensor streams. This makes MSP useful for examining low-level control over data fusion, performance, synchronization, and experimental logging.

\textbf{ReactGenie.}
ReactGenie is a development framework that uses large language models to generate complex multimodal interactions for React-based web interfaces \cite{Yang2024ReactGenie}. Developers specify multimodal scenarios, such as speech plus pointing on GUI components; ReactGenie then uses an LLM to synthesize interaction handlers and state management code, which can be reviewed and refined. ReactGenie represents a newer generation of LLM-assisted toolkits where part of the implementation logic is implicitly encoded in prompts, generated code, and model behavior rather than only in explicit APIs.

\textbf{WAMI.}
WAMI is an earlier toolkit for authoring multi-device, multimodal web applications \cite{Gruenstein2008WAMI}. It exposes a web-services interface to telephony applications and allows developers to define multimodal interaction flows that combine speech, DTMF input, and graphical interfaces. WAMI separates abstract multimodal interaction flows from concrete device bindings, allowing developers to target combinations of desktop, phone, and other devices. Compared to recent toolkits, it predates modern web standards and LLM-based approaches, but it remains a useful historical reference point.

\textbf{EmoSync.}
EmoSync is a recent backend framework for synchronizing multimodal emotion data and logs in real time for user studies and deployments \cite{Tong2025EmoSync}. It focuses on data collection, experiment control, and integration with multimodal large language models rather than UI widgets. EmoSync is particularly relevant for the benchmark dimension on experimental and logging support, because it treats temporal alignment and fine-grained emotion recognition as first-class concerns.

\subsection{Evaluating Toolkits in Computer Science and HCI}

Evaluating development kits is a common task in Computer Science and related disciplines. In the context of game development, Kilijanek and Miłosz compared Unity and Unreal Engine, two dominant game engines with different design philosophies, programming models, rendering pipelines, and production workflows \cite{HeroViredUnityUnreal}. This example is useful because it shows that toolkits and engines are not evaluated only by raw performance. They are also evaluated through usability, ecosystem support, workflow, documentation, extensibility, and the type of projects they enable.

HCI toolkit evaluation faces similar challenges. Ledo et al. reviewed 68 HCI toolkit papers and identified four main evaluation strategies: demonstrations, usage studies with developers, technical benchmarks, and heuristic evaluations \cite{Ledo2018ToolkitEval}. They argue that evaluations should match toolkit claims. If a toolkit claims better performance, technical benchmarks are needed. If it claims better usability or reduced developer effort, a developer study is appropriate. If it claims expressive power, demonstrations and comparative examples may be more suitable.

This insight is central for multimodal UI toolkits. Such toolkits often make several types of claims at once. A toolkit may claim to support more modalities, reduce implementation effort, enable faster prototyping, improve logging, support experiments, or integrate with modern AI methods. A single evaluation strategy is unlikely to capture all these claims. Therefore, a benchmark framework should separate structural capabilities, developer-facing workflow, and experimental support rather than collapsing them into one overall score.

\subsection{Benchmarking as a Method}

Benchmarking is a systematic method for evaluating and comparing systems, processes, or methods by defining clear criteria and comparing against reference standards or peers. In computational sciences, rigorous benchmarking studies are designed with explicit evaluation objectives, representative tasks, reproducible procedures, and unbiased metric comparison \cite{BioInfoBenchmarking2019}. These guidelines are useful beyond computational biology because they highlight general principles: define the evaluation target, choose representative tasks, report metrics transparently, and avoid overgeneralizing from narrow comparisons.

Benchmarking has also been discussed within quality management literature as a process for identifying improvement opportunities and guiding performance enhancement through structured steps such as planning, data collection, analysis, and action follow-up \cite{BhuttaHuq1999Benchmarking}. For multimodal UI toolkits, this means a benchmark should not merely list features. It should clarify what comparison is for, what evidence is being collected, and how future developers or researchers can use the comparison.

In multimodal UI development, benchmarking should not be limited to runtime performance. A toolkit may be technically powerful but difficult to configure. Another may be easy to use but limited in modality coverage. A third may provide excellent logging but weak GUI integration. A fourth may reduce coding effort through LLM-based generation while making reproducibility more difficult. Therefore, the benchmark proposed in this paper is designed to expose trade-offs rather than produce a single ranking.

% =========================================================
% 3. BENCHMARKING FRAMEWORK
% =========================================================
\section{Benchmarking Framework}

The proposed framework organizes multimodal UI toolkit evaluation into three high-level dimensions. Each dimension contains three criteria, resulting in nine benchmark criteria overall. The framework is designed to be reusable: researchers can apply it to the five toolkits discussed here, or extend it to additional research prototypes, commercial SDKs, augmented-reality frameworks, or accessibility-oriented multimodal systems.

\subsection{Toolkit Selection and Inclusion Criteria}

The initial benchmark focuses on Geno, MSP, ReactGenie, WAMI, and EmoSync. This set was chosen because it spans multiple eras and design philosophies. WAMI represents an older web-services architecture; Geno represents voice-based authoring on existing web applications; MSP represents modular multisensor pipelines; ReactGenie represents LLM-assisted multimodal interaction generation; and EmoSync represents emotion-centered multimodal logging and synchronization.

The inclusion criteria are:

\begin{itemize}
\item explicit support for at least two modalities, sensors, or multimodal data streams;
\item availability of a developer-facing API, framework, or tool;
\item documentation or publication material sufficient to understand the implementation model;
\item relevance for research, prototyping, or toolkit evaluation rather than only a closed commercial SDK.
\end{itemize}

This selection is not intended to cover the entire multimodal toolkit landscape. Instead, it provides a representative set for developing and illustrating the benchmark framework.

\subsection{B1: Modality Coverage and Interaction Abstraction}

The first dimension captures the structural capabilities of a toolkit. It asks what modalities are supported, how those modalities are represented, how they are combined, and how easily new modalities can be added.

\begin{itemize}
\item \textbf{Supported modalities}: speech, gesture, gaze, pen, touch, GUI context, biosignals, sensor streams, emotion data, or other inputs.
\item \textbf{Fusion abstraction}: pipeline-based processing, event-driven composition, service-based integration, model-centric fusion, or hybrid mechanisms.
\item \textbf{Extensibility mechanisms}: plugins, APIs, configurable nodes, external services, generated code, model prompts, or integration hooks.
\end{itemize}

This dimension is mainly evaluated through document analysis, API inspection, and prototype implementation. It is not primarily subjective, because supported modalities and fusion mechanisms can be observed in the toolkit architecture. However, practical validation is still useful because documentation may overstate how easy it is to use or extend these mechanisms in real development contexts.

\subsection{B2: Developer Experience and Workflow}

The second dimension captures the developer-facing experience of using the toolkit. It asks how easy it is to understand, configure, debug, and iterate with the toolkit.

\begin{itemize}
\item \textbf{Documentation clarity}: quality of tutorials, examples, API documentation, and conceptual explanations.
\item \textbf{Debugging support}: visibility of errors, logs, event streams, generated code, or live inspection tools.
\item \textbf{Workflow smoothness}: effort needed to move from setup to working prototype, including dependency management and iteration speed.
\end{itemize}

This dimension should be evaluated through a developer study or structured expert walkthrough. Unlike B1, it cannot be fully assessed from documentation alone because it depends on the developer's practical experience. A toolkit can expose a powerful API while still being difficult to configure or debug. Conversely, a toolkit can hide complexity and make initial prototyping faster while reducing transparency and low-level control.

\subsection{B3: Experimental and Integration Support}

The third dimension captures whether a toolkit supports research-oriented evaluation and integration into experimental workflows.

\begin{itemize}
\item \textbf{Logging facilities}: ability to record multimodal events, system states, user actions, model outputs, and timing data.
\item \textbf{Synchronization support}: ability to align sensor streams, interface events, speech, video, annotations, or model outputs in time.
\item \textbf{Scenario experiments}: support for replaying interactions, structuring trials, comparing configurations, or running controlled studies.
\end{itemize}

This dimension is especially important for HCI research. Many multimodal prototypes are not only built for deployment; they are also built for studies, experiments, and analysis. If a toolkit makes interaction easy but does not support logging or synchronization, researchers may still need substantial custom infrastructure. Conversely, a backend-oriented framework may provide strong experimental support while offering less help for GUI authoring.

\subsection{Framework Summary}

Table~\ref{tab:dimensions} summarizes the benchmark dimensions and criteria. The table is intentionally compact so that it remains readable in a one-column preprint format.

\begin{table}[H]
\centering
\caption{Benchmark dimensions and criteria for multimodal UI toolkit evaluation}
\label{tab:dimensions}
\small
\renewcommand{\arraystretch}{1.25}
\begin{tabularx}{\textwidth}{p{3.2cm}Y}
\toprule
\textbf{Dimension} & \textbf{Criteria and evaluation focus} \\
\midrule
\textbf{B1: Modality coverage and interaction abstraction} &
Supported modalities; fusion abstraction; extensibility mechanisms. This dimension captures what a toolkit structurally supports and how modalities are represented, combined, and extended. \\
\addlinespace
\textbf{B2: Developer experience and workflow} &
Documentation clarity; debugging support; workflow smoothness. This dimension captures how developers experience setup, implementation, debugging, and iteration. \\
\addlinespace
\textbf{B3: Experimental and integration support} &
Logging facilities; synchronization support; scenario experiments. This dimension captures whether the toolkit supports reproducible studies, event capture, stream alignment, and controlled evaluation. \\
\bottomrule
\end{tabularx}
\end{table}

% =========================================================
% 4. TOOLKIT COMPARISON
% =========================================================

\section{Illustrative Toolkit Comparison}

The framework can be instantiated by comparing the five selected toolkits across their main focus, example modalities, and experimental support. The following comparison is not presented as final empirical evidence. Instead, it illustrates how the framework structures comparison and prepares future empirical evaluation.

To avoid false precision, Table~\ref{tab:toolkitprofiles} uses one row per toolkit and describes each toolkit in a compact profile format. This is more appropriate for a framework paper than presenting numerical scores before the benchmark has been empirically instantiated. This profile-based presentation also keeps the comparison readable in a one-column preprint format, where dense multi-column tables can easily reduce clarity. It further emphasizes that the goal of this section is interpretive positioning rather than final scoring, leaving numerical ratings for future empirical benchmark studies. The profiles also preserve the original seminar paper's comparative logic while adapting it to a more publication-oriented layout. In this way, the section functions as a bridge between the conceptual benchmark dimensions and the more concrete evaluation protocol described in the next section.

\newpage
\subsection{Toolkit Profiles}

The following profiles summarize the five representative toolkits according to the proposed benchmark dimensions.

\begin{table}[H]
\centering
\caption{One-column comparison profile of five representative multimodal UI toolkits}
\label{tab:toolkitprofiles}
\small
\renewcommand{\arraystretch}{1.2}
\begin{tabularx}{\textwidth}{p{2.1cm}Y}
\toprule
\textbf{Toolkit} & \textbf{Profile according to the proposed benchmark} \\
\midrule
\textbf{Geno} &
\textbf{Main focus:} voice interaction on existing web applications. 
\textbf{Modalities:} speech plus GUI context, including mouse, keyboard, and DOM elements. 
\textbf{Experimental support:} useful command and GUI-event logging, but no full built-in study orchestration. Geno is strongest as a web-facing authoring tool for speech interaction. \\
\addlinespace
\textbf{MSP} &
\textbf{Main focus:} general multimodal and multisensor processing pipelines. 
\textbf{Modalities:} eye tracking, IMU, audio, video, and other sensor streams. 
\textbf{Experimental support:} Python logging and modular nodes, with flexible custom scripts for experiments. MSP is strongest for low-level pipeline control and sensor extensibility. \\
\addlinespace
\textbf{ReactGenie} &
\textbf{Main focus:} LLM-backed multimodal React applications. 
\textbf{Modalities:} speech plus touch or pointing on GUI components. 
\textbf{Experimental support:} standard web instrumentation and generated interaction logic, with analysis often left to external tools. ReactGenie is strongest for rapid prototyping but raises questions about generated-code transparency. \\
\addlinespace
\textbf{WAMI} &
\textbf{Main focus:} web-accessible, scenario-based multimodal interfaces. 
\textbf{Modalities:} speech, pen, mouse, GUI events, and telephony. 
\textbf{Experimental support:} scenario logs and replay, supporting controlled experiments through scripted scenarios. WAMI is strongest as a historical reference for abstract multimodal flow modeling. \\
\addlinespace
\textbf{EmoSync} &
\textbf{Main focus:} multimodal emotion-logging backend. 
\textbf{Modalities:} audio, video, text, facial expressions, speech, and transcripts. 
\textbf{Experimental support:} strong logging, replay, annotation, and synchronization across modalities. EmoSync is strongest for experiment-oriented multimodal data collection and synchronization. \\
\bottomrule
\end{tabularx}
\end{table}

The comparison highlights that multimodal UI toolkits should not be ranked through a single score. A toolkit may score highly on modality coverage but poorly on workflow smoothness. Another may provide strong logging but limited GUI integration. A third may reduce development effort through LLM assistance while making internal behavior harder to inspect.

\subsection{Benchmark Matrix Template}

A future empirical study can instantiate the framework as a matrix. However, a dense nine-criteria matrix is difficult to read in a one-column paper. Therefore, Table~\ref{tab:matrix} presents each toolkit across B1, B2, and B3 in a condensed format. Future work can replace these qualitative descriptors with measured values, such as setup time, task completion time, error rates, subjective workload, synchronization latency, or logging completeness.

\begin{table}[H]
\centering
\caption{Condensed B1--B3 benchmark matrix for future empirical comparison}
\label{tab:matrix}
\small
\renewcommand{\arraystretch}{1.2}
\begin{tabularx}{\textwidth}{p{2.1cm}Y}
\toprule
\textbf{Toolkit} & \textbf{B1--B3 interpretation} \\
\midrule
\textbf{Geno} &
\textbf{B1:} speech and GUI-context grounding; event-based web interaction. 
\textbf{B2:} strong for web demonstrations but dependent on documentation and browser setup. 
\textbf{B3:} useful command and event logs, but limited study orchestration. \\
\addlinespace
\textbf{MSP} &
\textbf{B1:} flexible multisensor pipelines and strong extensibility. 
\textbf{B2:} more setup effort, requiring comfort with Python and sensor configuration. 
\textbf{B3:} strong logging and custom experimental scripting. \\
\addlinespace
\textbf{ReactGenie} &
\textbf{B1:} LLM-assisted multimodal interaction generation for React. 
\textbf{B2:} potentially fast prototyping, but requires prompt steering and code review. 
\textbf{B3:} depends on external instrumentation and transparency of generated logic. \\
\addlinespace
\textbf{WAMI} &
\textbf{B1:} web-service and scenario-based multimodal flows. 
\textbf{B2:} older architecture, useful as a conceptual reference but less aligned with modern stacks. 
\textbf{B3:} scenario logs and replay support controlled interaction studies. \\
\addlinespace
\textbf{EmoSync} &
\textbf{B1:} emotion-centered multimodal data synchronization. 
\textbf{B2:} backend-oriented workflow, less focused on GUI authoring. 
\textbf{B3:} strong logging, annotation, and synchronization support. \\
\bottomrule
\end{tabularx}
\end{table}

This matrix is intended as a framework template rather than a completed result table. It makes the comparison readable in one-column format while preserving the three-dimensional logic of the original seminar draft.

% =========================================================
% 5. EVALUATION PROTOCOL
% =========================================================
\section{Proposed Evaluation Protocol}

The proposed framework can be instantiated through a mixed-method evaluation protocol. The protocol combines document and API analysis, prototype tasks, developer-study measures, and technical or experimental measures. This combination follows the principle that toolkit evaluations should match toolkit claims \cite{Ledo2018ToolkitEval}.

\subsection{Document and API Analysis}

The first step is document and API analysis. Researchers should collect official papers, documentation, tutorials, example projects, API references, and public repositories where available. For each toolkit, the analysis should identify supported modalities, fusion abstractions, extensibility mechanisms, logging features, synchronization support, and example workflows.

This step provides a reproducible baseline. Before developers are asked to evaluate a toolkit, researchers should understand what the toolkit officially claims to support and how those claims are represented in documentation. Document analysis also helps distinguish between a toolkit's conceptual design and its practical developer-facing workflow.

\subsection{Prototype Tasks}

The second step is to define small benchmark tasks that can be implemented across multiple toolkits. Example tasks may include:

\begin{itemize}
\item a voice-controlled web macro;
\item a speech-plus-pointing interaction on a GUI component;
\item a two-sensor logging pipeline;
\item a simple multimodal emotion-data synchronization task;
\item a scenario replay or trial-structured experimental setup.
\end{itemize}

The tasks should be realistic but limited in scope. They should be based on official examples where possible, then slightly extended to test whether the toolkit supports meaningful adaptation. This avoids evaluating only polished demo paths while still keeping the benchmark manageable.

\subsection{Developer Study Design}

A future empirical instantiation of the framework could use a within-subjects developer study with approximately \(N \approx 20{-}30\) participants who have basic web or Python development experience. Each participant would implement two short tasks using two different toolkits, counter-balanced across participants \cite{Ledo2018ToolkitEval}.

The study could collect task completion time, number of errors or restarts, perceived ease of use, perceived control, documentation clarity, debugging support, and qualitative comments. These measures would directly instantiate the B2 dimension while also revealing whether the B1 and B3 capabilities identified in document analysis are usable in practice.

The goal of the developer study would not be to produce a universal ranking of all multimodal UI toolkits. Instead, it would test whether the proposed framework captures meaningful differences in workflow, abstraction level, and experimental support.

\subsection{Technical and Experimental Measures}

Technical measures should be selected according to toolkit claims. Pipeline-based toolkits may be evaluated through latency, synchronization accuracy, throughput, or logging completeness. LLM-assisted toolkits may be evaluated through generated-code correctness, transparency, edit effort, and reproducibility. Experiment-oriented toolkits may be evaluated through scenario replay, annotation support, export formats, and the ability to reconstruct interaction sequences after a study session.

The choice of technical metrics should also reflect the type of multimodal data being handled. For sensor-heavy systems, timing precision and synchronization stability are central because even small temporal misalignments can affect interpretation of gaze, gesture, audio, or physiological signals. For web-based authoring tools, the relevant measures may instead include setup time, number of manual steps, generated code size, and robustness across interface changes. For emotion-aware or AI-supported systems, additional measures should include traceability of model outputs, consistency across repeated runs, and the degree to which developers can inspect or override automated decisions.

This flexible approach prevents unfair comparison. A voice-command authoring tool, a sensor pipeline, and an emotion-logging backend do not all optimize for the same outcome. The benchmark therefore compares them across shared dimensions while allowing toolkit-specific evidence. In practice, this means that each toolkit should be evaluated against both common benchmark dimensions and claim-specific metrics. Such a structure allows the benchmark to remain comparable across systems while still respecting the different design goals, technical architectures, and intended use cases of multimodal UI toolkits.

% =========================================================
% 6. VISUALIZATION
% =========================================================
\newpage
\section{Visualizing Benchmark Results}

\begin{figure}[H]
\centering
\includegraphics[width=0.92\textwidth]{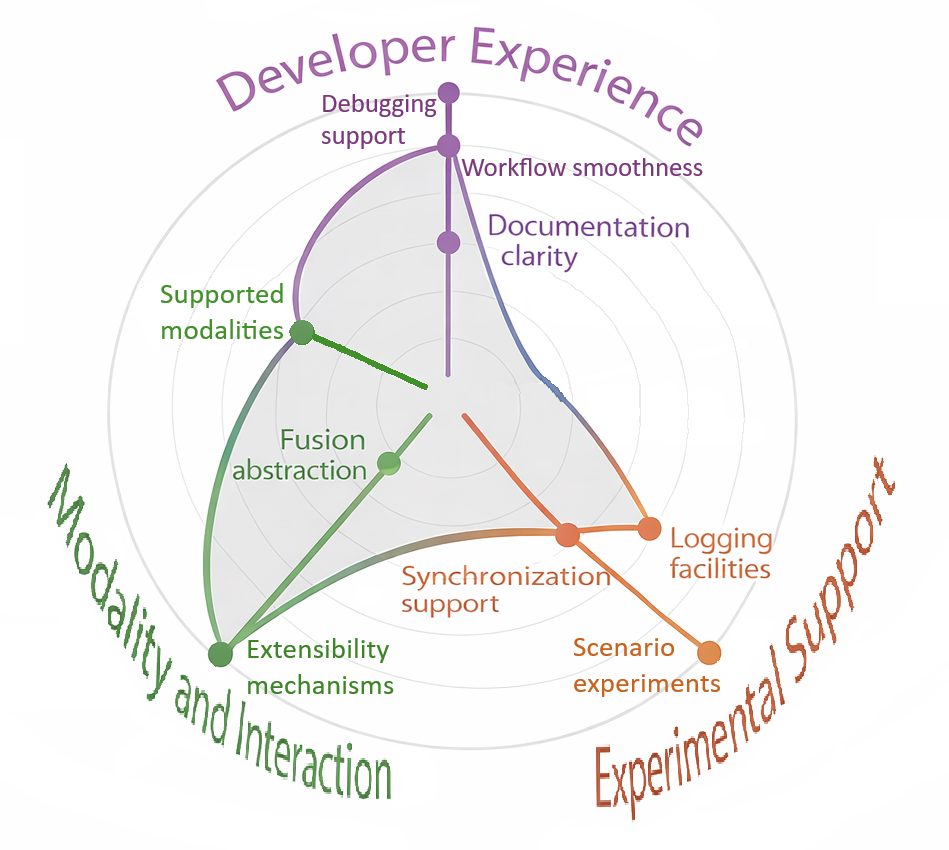}
\caption{Illustrative visualization concept for benchmark results across modality coverage, developer workflow, and experimental support.}
\label{fig:benchmark-visual}
\end{figure}

Benchmark results should be presented through a combination of tables and charts. Tables are useful for describing supported modalities, API features, documentation quality, and experimental support. Charts are useful for communicating trade-offs across dimensions, as illustrated in Figure~\ref{fig:benchmark-visual}.

The figure is intentionally illustrative. In a completed empirical study, the axes would be populated by measured or coded values. B1 values could be derived from document analysis, B2 values from developer-study ratings, and B3 values from logging and synchronization tests.

Such visualizations should not be used to hide methodological complexity. Instead, they should make trade-offs visible while accompanying tables explain how each score or label was produced. For example, a toolkit might appear strong in developer workflow but weak in experimental support, or strong in synchronization but weaker in rapid GUI prototyping.

% =========================================================
% 7. DISCUSSION
% =========================================================
\newpage
\section{Discussion}

\subsection{Why Multimodal Toolkit Benchmarking Is Difficult}

Benchmarking multimodal UI toolkits is difficult because the toolkits do not all solve the same problem. Geno supports voice-based authoring on existing web applications, MSP supports flexible sensor pipelines, ReactGenie supports LLM-assisted React interaction generation, WAMI supports web-accessible multimodal interaction flows, and EmoSync supports emotion-focused logging and synchronization \cite{Sarmah2020Geno,Barz2021MSP,Yang2024ReactGenie,Gruenstein2008WAMI,Tong2025EmoSync}.

This diversity means that benchmark design must avoid false equivalence. A toolkit with strong low-level control may require more setup effort. A toolkit with high-level automation may reduce initial effort but make debugging harder. A toolkit with strong logging may be less suitable for rapid GUI prototyping. A toolkit with strong AI support may accelerate development but introduce new concerns around reproducibility and transparency.

The proposed framework addresses this by comparing toolkits across multiple dimensions rather than forcing a single overall ranking. It also separates observable system capabilities from developer-facing experience and experimental support. This separation is important because a toolkit may be structurally powerful but practically difficult, or easy to prototype with but weaker for controlled studies.

\subsection{The Role of LLM-Based Toolkits}

ReactGenie represents a broader shift toward LLM-assisted development. In such systems, developers may write less explicit code, but they must evaluate generated interaction logic, inspect model-produced handlers, and decide whether the generated behavior is reproducible and maintainable \cite{Yang2024ReactGenie}.

This changes the meaning of developer effort. Effort is no longer only measured by lines of code or implementation time. It also includes prompt formulation, review, debugging, correction, and trust calibration. Future benchmarks for multimodal UI toolkits should therefore include criteria that capture transparency and controllability in AI-assisted workflows.

LLM-based toolkits also raise methodological challenges for benchmarking. If generated code changes across model versions, prompts, or hidden system updates, reproducibility becomes more difficult. A benchmark involving LLM-based multimodal toolkits should therefore document model versions, prompts, generated code, human edits, and runtime environments. Without this documentation, two researchers may believe they are evaluating the same system while actually using different generated interaction logic.

\subsection{Benchmarking as a Framework Contribution}

The main contribution of this paper is not a final empirical ranking of five toolkits. Instead, it is a framework that can guide future empirical comparison. This is important because many HCI toolkit papers present strong demonstrations but limited systematic evaluation \cite{Ledo2018ToolkitEval}.

A reusable benchmark framework can help researchers design fairer comparisons, report clearer evidence, and identify where toolkits support or fail real development workflows. It can also help developers choose toolkits based on project needs rather than general popularity or isolated demonstrations.

The framework is especially useful when toolkits differ in purpose. Geno, MSP, ReactGenie, WAMI, and EmoSync are not interchangeable products. They represent different approaches to multimodal development. A fair comparison should therefore make their trade-offs visible rather than treating them as competitors in a single linear ranking.

% =========================================================
% 8. LIMITATIONS
% =========================================================
\newpage
\section{Limitations}

This paper is a benchmarking framework paper, not a completed empirical benchmark. The toolkit comparisons are illustrative and based on literature analysis, toolkit descriptions, and framework reasoning. They should not be interpreted as final measured results.

Second, the framework currently focuses on five representative toolkits. Although these systems cover different eras and design philosophies, the multimodal toolkit landscape is broader. Future work should include additional research prototypes, commercial SDKs, augmented-reality toolkits, accessibility-oriented toolkits, and mobile multimodal frameworks.

Third, the proposed developer study has not yet been conducted. Therefore, claims about developer experience, workflow smoothness, and debugging support should be treated as proposed evaluation dimensions rather than completed findings. The framework identifies what should be measured, but future studies must still collect empirical evidence.

Finally, LLM-based toolkits introduce rapid version changes. A benchmark involving AI-assisted systems may become outdated more quickly than a benchmark involving static libraries. Future benchmark protocols should therefore document toolkit versions, model versions, prompts, dependencies, and runtime environments.

% =========================================================
% 9. FUTURE WORK
% =========================================================
\section{Future Work}

Future work should instantiate the proposed benchmark with empirical measurements. A first step would be a small developer study with 10--20 participants implementing controlled tasks across two or three selected toolkits. This would provide initial evidence for documentation clarity, debugging support, workflow smoothness, and perceived control.

A second step would be technical benchmarking of selected toolkit features. MSP and EmoSync could be evaluated through logging completeness, synchronization accuracy, latency, or replay support. Geno and ReactGenie could be evaluated through implementation time, generated-code quality, correctness of multimodal handlers, and developer correction effort.

A third step would be to extend the framework beyond the five toolkits discussed here. The benchmark could include augmented-reality systems, accessibility-focused multimodal interfaces, commercial SDKs, and future LLM-backed development environments.

Finally, future work should investigate reproducibility in LLM-assisted multimodal UI development. When interaction logic is generated or influenced by a model, researchers must document prompts, versions, generated code, and human edits. This will become increasingly important as multimodal interface development shifts from explicit programming toward AI-supported authoring.

% =========================================================
% 10. CONCLUSION
% =========================================================
\section{Conclusion}

This paper proposed a benchmarking framework for multimodal user interface toolkits. While multimodal interaction is becoming increasingly central to modern interfaces, developers still face fragmented tooling and inconsistent support for modality fusion, synchronization, and logging, making systematic comparison both necessary and challenging.

By synthesizing insights from prior toolkit evaluations and benchmarking literature, the paper proposed a structured set of benchmark dimensions and a mixed-method evaluation strategy that combines document analysis, technical comparison, and a future developer study. This approach is intended to surface not only which features are supported, but also how design choices shape developer effort and workflow in practice.

Using Geno, MSP, ReactGenie, WAMI, and EmoSync as representative examples, the paper showed that multimodal toolkits differ not only in supported modalities but also in abstraction level, developer workflow, logging support, synchronization mechanisms, and relationship to modern AI-assisted development. These differences make single-score comparisons insufficient.

The proposed framework therefore provides a foundation for future empirical work and clearer decision-making in multimodal UI research. It can help researchers compare toolkits more fairly, help developers choose tools more deliberately, and help the field move toward more transparent evaluation practices for multimodal user interface development.

% =========================================================
% REFERENCES
% =========================================================


{\footnotesize
\begin{thebibliography}{10}

\bibitem{Barz2021MSP}
Michael Barz, Omair Shahzad Bhatti, Bengt Lüers, Alexander Prange, and
Daniel Sonntag. 2021.
\newblock Multisensor-Pipeline: A Lightweight, Flexible, and Extensible
Framework for Building Multimodal-Multisensor Interfaces.
\newblock In {\em Companion Publication of the 2021 International Conference
on Multimodal Interaction (ICMI '21 Companion)}.
\newblock \url{https://doi.org/10.1145/3461615.3485432}.

\bibitem{Sarmah2020Geno}
Ritam Jyoti Sarmah, Yunpeng Ding, Di Wang, Cheuk Yin Phipson Lee,
Toby Jia-Jun Li, and Xiang ``Anthony'' Chen. 2020.
\newblock Geno: A Developer Tool for Authoring Multimodal Interaction on
Existing Web Applications.
\newblock In {\em Proceedings of the 33rd Annual ACM Symposium on User
Interface Software and Technology (UIST '20)}.
\newblock \url{https://doi.org/10.1145/3379337.3415848}.

\bibitem{Yang2024ReactGenie}
Jackie Junrui Yang, Yingtian Shi, Yuhan Zhang, Karina Li, Daniel Wan Rosli,
Anisha Jain, Shuning Zhang, Tianshi Li, James A. Landay, and Monica S. Lam.
2024.
\newblock ReactGenie: A Development Framework for Complex Multimodal
Interactions Using Large Language Models.
\newblock In {\em Proceedings of the 2024 CHI Conference on Human Factors in
Computing Systems (CHI '24)}.
\newblock \url{https://doi.org/10.1145/3613904.3642517}.

\bibitem{Gruenstein2008WAMI}
Alexander Gruenstein, Ian McGraw, and Ibrahim Badr. 2008.
\newblock The WAMI toolkit for developing, deploying, and evaluating
Web-Accessible multimodal interfaces.
\newblock In {\em Proceedings of the 10th International Conference on
Multimodal Interfaces (ICMI '08)}.
\newblock \url{https://doi.org/10.1145/1452392.1452420}.

\bibitem{Tong2025EmoSync}
Jintao Tong, Shiwei Li, Zijian Zhuang, Jinghan Hu, and Yixiong Zou. 2025.
\newblock EmoSync: Multi-Stage Reasoning with Multimodal Large Language Models
for Fine-Grained Emotion Recognition.
\newblock In {\em Proceedings of the 3rd International Workshop on Multimodal
and Responsible Affective Computing (MRAC '25)}.
\newblock \url{https://doi.org/10.1145/3746270.3760231}.

\bibitem{Lee2024GazePointAR}
Thibaut Septon, Santiago Villarreal-Narvaez, Xavier Devroey, and
Bruno Dumas. 2024.
\newblock Exploiting Semantic Search and Object-Oriented Programming to Ease
Multimodal Interface Development.
\newblock In {\em Proceedings of the 16th ACM SIGCHI Symposium on Engineering
Interactive Computing Systems (EICS '24)}.
\newblock \url{https://doi.org/10.1145/3660515.3664244}.

\bibitem{Ledo2018ToolkitEval}
David Ledo, Steven Houben, Jo Vermeulen, Nicolai Marquardt, Lora Oehlberg, and
Saul Greenberg. 2018.
\newblock Evaluation Strategies for HCI Toolkit Research.
\newblock In {\em Proceedings of the 2018 CHI Conference on Human Factors in
Computing Systems (CHI '18)}.
\newblock \url{https://doi.org/10.1145/3173574.3173610}.

\bibitem{BioInfoBenchmarking2019}
Lukas M. Weber, Wouter Saelens, Robrecht Cannoodt, Charlotte Soneson,
Alexander Hapfelmeier, Paul P. Gardner, Anne-Laure Boulesteix, Yvan Saeys,
and Mark D. Robinson. 2019.
\newblock Essential guidelines for computational method benchmarking.
\newblock \emph{Genome Biology} 20, 125.
\newblock Retrieved from
\newblock \url{https://genomebiology.biomedcentral.com/articles/10.1186/s13059-019-1738-8}.

\bibitem{BhuttaHuq1999Benchmarking}
R. Dattakumar and R. Jagadeesh. 2003.
\newblock A review of literature on benchmarking.
\newblock \emph{Benchmarking: An International Journal} 10, 3 (June 2003), 176--209.
\newblock Retrieved from
\newblock \url{https://www.researchgate.net/publication/235312564_A_review_of_literature_on_benchmarking}.

\bibitem{HeroViredUnityUnreal}
Robert Kilijanek and Marek Miłosz. 2025.
\newblock Comparative analysis of the performance of Unity and Unreal Engine.
\newblock {\em Journal of Computer Sciences Institute} 35, 197--201.
\newblock \url{https://doi.org/10.35784/jcsi.7298}.

\bibitem{Metatla2019VoiceUISchools}
Oussama Metatla, Alison Oldfield, Taimur Ahmed, Antonis Vafeas, and Sunny Miglani. 2019.
\newblock Voice User Interfaces in Schools: Co-designing for Inclusion with Visually-Impaired and Sighted Pupils.
\newblock In {\em Proceedings of the 2019 CHI Conference on Human Factors in Computing Systems (CHI '19)}.
\newblock \url{https://doi.org/10.1145/3290605.3300608}.

\end{thebibliography}
}
\end{document}